\documentclass[lettersize,journal]{IEEEtran}

\usepackage{amsmath,amsfonts}
\usepackage{algorithmicx}
\usepackage{algorithm}
\usepackage{algpseudocode}
\usepackage{array}
\usepackage[caption=false,font=normalsize,labelfont=sf,textfont=sf]{subfig}
\usepackage{lettrine}
\usepackage{textcomp}
\usepackage{stfloats}
\usepackage{url}
\usepackage{verbatim}
\usepackage{graphicx}
\usepackage{subcaption} 
\usepackage{cite}
\usepackage{flushend}
\usepackage{float}
\usepackage{amsthm}
\usepackage{cuted}
\usepackage{amssymb}
\usepackage{tikz}
\usepackage{multirow}
\usepackage{mathtools, cuted}
\usepackage{lscape}
\usepackage[numbers,sort&compress]{natbib}
\usepackage{algpseudocode}
\usepackage{comment}
\usepackage{titlesec}
\usepackage{subcaption}
\usepackage{mathrsfs}
\IEEEoverridecommandlockouts
\begin{document}

\flushend

\title{Optimization of Joint Beamforming in RIS-aided Multi-carrier and Multi-user MIMO Systems}
\author{Nuwan Balasuriya, Atefeh Termehchi, and Ekram Hossain, {\em Fellow, IEEE}}
\renewcommand\footnotemark{}
\thanks{The authors are with the Department of Electrical and Computer Engineering, University of Manitoba, Winnipeg, MB, Canada (emails: umbalasd@myumanitoba.ca, \{Atefeh.Termehchi, Ekram.Hossain\}@umanitoba.ca).}

\maketitle 
\renewcommand{\IEEEaftertitletext}{\vspace{-20mm}}

\begin{abstract}
Reconfigurable intelligent surface (RIS)-aided multiple-input multiple-output (MIMO) communication is a promising enabler for beyond–fifth-generation wireless networks. However, RIS phase shifters typically rely on a tuning element (for example a variable-length stub), which can be optimally configured for only one carrier frequency at a time. This constraint complicates RIS deployment in multi-carrier systems and often necessitates additional measures, such as using multiple RIS panels or partitioning RIS elements across carriers.
In this work, we propose a novel scheme in which all RIS elements simultaneously serve multiple carriers, delivering improved performance over existing approaches. Our method leverages the multiple reflections captured by physically consistent RIS models to realize the desired phase shifts for all carriers using a single stub length setting. We further develop a joint active and passive beamforming optimization framework tailored to multi-carrier, physically consistent RIS-aided systems. Simulation results confirm the effectiveness of the proposed model and demonstrate clear gains over conventional models.
\end{abstract}


\section{Introduction}\label{Intro}
\lettrine{R}{econfigurable} intelligent surfaces (RIS) have recently attracted considerable attention as a promising means to mitigate shadowing and fading in wireless communication systems. As largely passive devices, RIS impose controllable phase shifts on the multipath channel, shaping signal propagation so that receivers can obtain stronger and more reliable links. This capability can improve coverage and increase system capacity. In particular, RIS-aided multiple-input multiple-output (MIMO) systems have emerged as a key enabler for beyond–fifth-generation (5G) wireless networks, motivating extensive work on optimizing RIS-assisted MIMO transmissions \cite{Daj22, Has24}. More recently, RIS optimization has also been studied under physically consistent models. For instance, \cite{Wij24} considers multiple-reflection effects induced by inter-element mutual coupling, \cite{Che20} investigates the influence of the angle of incidence, and \cite{Cao24} examines practical RIS phase-shifter limitations arising from constrained phase tunability.

Multi-carrier communication adds another degree of freedom by enabling diversity and/or multiplexing gains across carriers. RIS-aided multi-carrier MIMO systems therefore have the potential to combine the benefits of RIS with those of multi-carrier techniques. A key challenge, however, is that RIS phase shifts are commonly controlled by a mechanism such as a varactor diode capacitance or a phase shifter stub length. Because the resulting phase shift depends jointly on the above parameter value and the carrier frequency, a fixed RIS configuration generally produces different phase responses across different carriers. On the other hand the optimum phase shift requirement at different carriers are different and cannot be simultaneously satisfied by selecting the possible single parameter value arrangement.

A simple workaround is to deploy a dedicated RIS panel for each carrier, but this is clearly impractical. Another option is to choose parameter values that compromise across carriers. However, this yields suboptimal phase shifts at each frequency and prevents full exploitation of the RIS. A different approach explored in \cite{Cai22, Sol23}, partitions the RIS so that only a subset of elements serves each carrier. While this enables accurate per-carrier phase control, it reduces the number of active elements per carrier, effectively lowering the channel rank and degrading capacity. Active RIS designs, such as those studied in \cite{Tua24}, can offer stronger beamforming flexibility but incur substantial power consumption, limiting their practicality. 

Physically-consistent RIS models explicitly capture the multiple-reflection phenomena that arise in densely packed arrays, opening new avenues for multi-carrier MIMO optimization using a single RIS configuration. In this work, we leverage such models and propose a joint beamforming optimization framework for RIS-aided multi-carrier MIMO systems. We benchmark our approach against existing strategies, including element-partitioning methods and active RIS architectures as in \cite{Tua24}, and demonstrate the advantages of the proposed framework.

\section{System Model and Assumptions}\label{sysmod}

We consider a single base station (BS) antenna array with $N$ elements serving $K (\leq N)$ single antenna user devices. A passive rectangular RIS array located close to the user devices consists of $L_H\times L_V=L (\geq K)$ individually configurable elements. BS transmits using $M$ equally spaced independent carriers. Let the downlink channel from BS to RIS in $m-$th carrier be $\textbf{H}_{bs,m} \in \mathbb{C}^{L \times N}$ and the channel from RIS to user $k$ be $\textbf{h}_{su,m,k} \in \mathbb{C}^{L}$. The direct channel from BS to user $k$ is given by $\textbf{h}_{bu,m,k} \in \mathbb{C}^{N}$. $\mathbf{\Upsilon}_m=\mbox{diag}(\textbf{v}_m)$, where $\textbf{v}_m \in \mathbb{C}^{L}$ represents the phase shift matrix of the RIS in $m$-th carrier. For the physically consistent model we consider the full phase shift matrix given by $\mathbf{\Phi}_m=\left(\mathbf{\Upsilon}^{-1}_m-\mathbf{S}_m\right)^{-1}$, where $\textbf{S}_m \in \mathbb{C}^{L \times L}$ represents the scattering matrix in the $m$-th carrier \cite{Mez22}.

\subsection{Scattering Matrix}
The scattering matrix $\mathbf{S}_m$ captures inter-element mutual coupling and multiple reflections, following the physically consistent model in [9]. Although individual RIS phase-shift coefficients have unit magnitude, the effective response $\boldsymbol{\Phi}_m$ is not restricted to element-wise unit magnitudes due to multiple reflections. When $\mathbf{S}_m=\mathbf{0}$, the model reduces to the conventional RIS phase-shift matrix $\boldsymbol{\Upsilon}_m$. With $\mathbf{S}_m\neq\mathbf{0}$, multiple reflections add path delays across elements, so the phases at different carriers are no longer locked by the same $\boldsymbol{\ell}$.
The scattering matrix is related to embedded pattern coupling matrix $\textbf{B}_m$ given by: 
\begin{align}
    &\left[\textbf{B}_m\right]_{w+L_V(x-1) , y+L_V(z-1)}= \nonumber \\ 
    &\quad \begin{cases} 
        \frac{\pi a^2}{\lambda_m^2}, & \text{if } w=y \text{ and } x=z, \\ 
        \frac{a}{\lambda_m} \times \frac{J_1\left(\frac{2\pi a}{\lambda_m} \sqrt{(w-y)^2+(x-z)^2}\right)}
        {\sqrt{(w-y)^2+(x-z)^2}}, & \text{otherwise}
    \end{cases}
\end{align}
where $a$ and $\lambda_m$ are the spacing between the RIS elements and wavelength of operation respectively \cite{Mez22}. $J_1(.)$ stands for the first order Bessel function of the first kind and $\left[\textbf{B}_m\right]_{w+L_V(x-1) , y+L_V(z-1)}$ is the embedded pattern coupling between the RIS elements located at indices $(w,x)$ and $(y,z)$. Note that $\textbf{B}_m$ is a real symmetric matrix. Hence, let $\textbf{U}_m\mathbf{\Lambda_m} \textbf{U}_m^H$ be the eigen value decomposition of $\textbf{B}_m$ with $\textbf{U}_m$ being a unitary matrix and $\mathbf{\Lambda_m}$ is a diagonal matrix having eigen values of $\textbf{B}_m$. Then, $\textbf{S}_m=\textbf{U}_m\sqrt{\textbf{I}_L-\mathbf{\lambda_m}}\textbf{U}_m^T$. Note that $\textbf{S}_m$ is a function of carrier frequency and element spacing, hence we denote $\textbf{S}_m=f(f_m,a)$.

\subsection{Phase Shifter Arrangement}
We consider stub-based phase shifters, capable of providing a full $[0,2\pi]$ phase range, with one phase shifter per RIS element. A stub is a transmission-line segment connected in series or parallel with the main line. Two arrangements are considered: (1) a mechanically adjustable stub length using a sliding arm, connected across the RIS antenna element, and (2) switch-selected discrete stub lengths, resulting in quantized phase shifts. Typically implemented as microstrip lines, stubs exhibit frequency-periodic phase responses, enabling phase variation over a wide bandwidth. The input impedance of an open-circuited stub is given by:
\begin{equation}
Z_{p,m}=-jZ_0 \cot\left(\beta_m \ell_p\right),
\end{equation}
where $\beta_m=\frac{2 \pi}{c}f_m$ and $Z_0, c, \ell_p$ are the characteristic impedance of the line, speed of radio frequency (RF) propagation and the $p-$th stub length respectively. Thus we can calculate the phase shift at the $p-$th element as:
\begin{equation}\label{eq28}
    v_{pp,m}=\frac{Z_{p,m}-Z_0}{Z_{p,m}+Z_0} 
    =\frac{-j\cot(\beta_m \ell_p)-1}{-j\cot(\beta_m \ell_p)+1}=e^{-2j\beta_m \ell_p}.
\end{equation}
Note that the resulting phase shift is a periodic function of frequency.

We first consider a continuous stub length and later, we will consider an available discrete stub length set arrangement.

\subsection{Signal Model}
We consider a transmit signal vector $\textbf{s}_m=[s_{m,1}, s_{m,2},\dots,s_{m,K}]^T$ precoded using an active precoder matrix $\textbf{W}_m=[\textbf{w}_{m,1}, \textbf{w}_{m,2},\dots, \textbf{w}_{m,K}]$ to produce a signal vector $\textbf{x}_m=[x_{m,1}, x_{m,2},\dots,x_{m,N}]^T$. This signal vector is then transmitted using $N$ BS antennas. From the BS to receiver mobiles, the signal takes two paths: one direct path and the other the reflected path via the RIS. The received signal vector at user $k$ is given by
\begin{eqnarray}
y_{m,k}=\alpha\left(\textbf{h}^H_{su,m,k}(\mathbf{\Upsilon}_m^{-1}-\textbf{S}_m)^{-1} \textbf{H}_{bs,m}\sum_{k'=1}^K \textbf{w}_{m,k'}\textbf{s}_{m,k'}\right.\\ \nonumber
    \left.+\textbf{h}^H_{bu,m,k}\sum_{k'=1}^K \textbf{w}_{m,k'}\textbf{s}_{m,k'}+n_{m,k}\right),
\end{eqnarray}
where $n_{m,k}$ is the additive white Gaussian noise (AWGN) at the receiver $k$ in carrier $m$ and $\alpha$ is a common scaling factor applied at each of the receivers.
Hence the overall receive signal vector is 
\begin{eqnarray}
\mathbf{y}_m &=& [y_{m,1}, y_{m,2},\dots,y_{m,K}]^T
= \alpha \mathbf{H}^H_m\mathbf{W}_m\mathbf{s}_m+\alpha\mathbf{n}_m,  \notag
\end{eqnarray}
where $\mathbf{H}^H_m=\mathbf{H}^H_{su,m}(\mathbf{\Upsilon}^{-1}_m-\mathbf{S}_m)^{-1}\mathbf{H}_{bs,m}
   + \mathbf{H}^H_{bu,m}$, $\textbf{H}_{su,m}=[\textbf{h}_{su,m,1}, \dots,\textbf{h}_{su,m,K}]$ and $\textbf{H}_{bu,m}=[\textbf{h}_{bu,m,1}, \dots, \textbf{h}_{bu,m,K}]$. BS has a total power budget of $\mathcal{P}_T$ Watts.

\section{Joint Beamforming optimization}\label{jointOpt}
Our objective is to enhance the overall sum-rate of the multi-carrier RIS-aided MIMO system while promoting fairness among users. Since direct sum-rate maximization can favor users with stronger channel conditions, potentially leading to unfair resource allocation, we adopt the sum-minimum mean square error criterion $\sum_{m=1}^M\mathbb{E}\left\{\|\textbf{y}_m-\textbf{s}_m\|^2_2\right\}$  \cite{Wij24, Reh21}.
As shown in \cite{Wij24}, this results in  the following optimization problem:
\begin{subequations}\label{first:main}
\begin{equation}\tag{\ref{first:main}}
 \min \limits_{\alpha, \{\textbf{W}_m\}, \bf{\ell}} \sum_{m=1}^M\|\alpha \textbf{H}^H_m \textbf{W}_m-\textbf{I}_K\|^2_F+KM\alpha^2 \sigma^2_n
\end{equation}
\begin{equation}\label{first:a}
\mbox{s.t.}~~~~~~\sum_{m=1}^M\|\textbf{W}_m\|^2_F=\mathcal{P}_T,
\end{equation}
\begin{equation}\label{first:a1}
\textbf{H}_m=\left[\textbf{H}^H_{su,m}(\mathbf{\Upsilon}^{-1}_m-\textbf{S}_m)^{-1}\textbf{H}_{bs,m}+\textbf{H}^H_{bu,m}\right]^H,
\end{equation}
\begin{equation}\label{first:a3}
\mbox{diag}(\mathbf{\Upsilon}_m)=[v_{11,m},\dots,v_{LL,m}]~\mbox{and}~v_{pp,m}=e^{-2j\beta_m \ell_p},  
\end{equation}
 \begin{equation}\label{first:a2}
v_{pq,m}=0 \mbox{ for } p\neq q,
\end{equation}
 \begin{equation}\label{first:a2}
0\leq\ell_p \leq \ell_{\mbox{max}}.
\end{equation}

\end{subequations}
This problem can then be converted to an alternating optimization between two sub-optimization problems:
\begin{enumerate}
    \item Optimization for $\alpha$ and $\{\textbf{W}_m\}_{m=1}^M$ under the constraint (\ref{first:a}) and known $\{\textbf{H}_m\}_{m=1}^M$, named as active beamforming optimization
    \item Optimization for $\bf{\ell}$ under the constraints (\ref{first:a1})-(\ref{first:a2}) and known $\alpha$ and $\{\textbf{W}_m\}_{m=1}^M$, named as passive beamforming optimization.
\end{enumerate}

\subsection{Active Beamforming Optimization}
The sub-problem of jointly optimizing $\alpha$ and $\{\textbf{W}_m\}_{m=1}^M$,
\begin{subequations}\label{second:main}
\begin{equation}\tag{\ref{second:main}}
 \min \limits_{\alpha, \{\textbf{W}_m\}} \sum_{m=1}^M\|\alpha \textbf{H}^H_m \textbf{W}_m-\textbf{I}_K\|^2_F+KM\alpha^2 \sigma^2_n
\end{equation}
\begin{equation}\label{second:a}\nonumber
~\mbox{s.t.}~~\sum_{m=1}^M\|\textbf{W}_m\|^2_F=\mathcal{P}_T
\end{equation}
\end{subequations}
can be equivalently reformulated as a convex quadratic optimization problem, which admits a closed-form solution.\\
\\
{\em \textbf{Result 1:}} 
The optimum solutions of (\ref{second:main}) are given by:
\begin{equation}\label{soln1}
\alpha=\frac{1}{\sqrt{{\mathcal{P}_T}}}\sqrt{\sum_{m=1}^M tr\left[\left(\textbf{H}_m\textbf{H}_m^H+\frac{KM\sigma_n^2}{\mathcal{P}_T}\textbf{I}_N\right)^{-2}\textbf{H}_m\textbf{H}_m^H\right]}
\end{equation}
and
\begin{equation}\label{soln2}
    \textbf{W}_m=\frac{1}{\alpha}\left(\textbf{H}_m\textbf{H}_m^H+\frac{KM\sigma_n^2}{{\mathcal{P}_T}}\textbf{I}_N\right)^{-1}\textbf{H}_m
\end{equation}
 for $m\in\{1,\dots,M\}$.
\\

{\em \textbf{Proof:}} Let us take $\alpha \textbf{W}_m=\textbf{W}'_m$. Hence, 
\begin{eqnarray}
    \alpha^2 \textbf{W}_m\textbf{W}^H_m&=&\textbf{W}'_m\textbf{W}'^H_m \\ 
     \alpha^2 tr\left(\textbf{W}_m\textbf{W}^H_m\right)&=&tr\left(\textbf{W}'_m\textbf{W}'^H_m\right) \\
     \alpha^2 \sum_{m=1}^M\|\textbf{W}_m\|^2_F&=&\sum_{m=1}^M\|\textbf{W}'_m\|^2_F \\ 
    \alpha^2 \mathcal{P}_T&=&\sum_{m=1}^M\|\textbf{W}'_m\|^2_F \label{rel1}
\end{eqnarray}
Applying (\ref{rel1}) in (\ref{second:main}) we obtain the modified sub-optimization problem:
\begin{equation}\label{newprob}
    \min \limits_{\{\textbf{W}'_m\}} \sum_{m=1}^M\|\textbf{H}^H_m \textbf{W}'_m-\textbf{I}_K\|^2_F+\frac{KM \sigma^2_n}{\mathcal{P}_T}\sum_{m=1}^M\|\textbf{W}'_m\|^2_F.
\end{equation}
Equating the first derivative of the objective function in (\ref{newprob}) to zero we obtain,
\begin{equation}
    \textbf{H}_m\left(\textbf{H}_m^H\textbf{W}'_m-\textbf{I}_K\right)+\frac{KM\sigma_n^2}{\mathcal{P}_T}\textbf{W}'_m=\textbf{0},~~\mbox{for}~~m\in\{1,\dots,M\}
\end{equation}
yielding  \begin{equation}\label{res2}
   \textbf{W}'_m=\left(\textbf{H}_m\textbf{H}_m^H+\frac{KM \sigma^2_n}{\mathcal{P}_T}\textbf{I}_N\right)^{-1}\textbf{H}_m.
\end{equation}
    
    Thus, $\textbf{W}_m=\frac{1}{\alpha}\left(\textbf{H}_m\textbf{H}_m^H+\frac{KM \sigma^2_n}{\mathcal{P}_T}\textbf{I}_N\right)^{-1}\textbf{H}_m$ for $m\in\{1,\dots,M\}$, as in (\ref{soln2}).

Using result (\ref{res2}) in (\ref{rel1}),
\begin{eqnarray}
\alpha^2 \mathcal{P}_T&=&\sum_{m=1}^M\left\lVert\left(\textbf{H}_m\textbf{H}_m^H+\frac{KM \sigma^2_n}{\mathcal{P}_T}\textbf{I}_N\right)^{-1}\textbf{H}_m\right \rVert^2_F \\ \nonumber
&=&\sum_{m=1}^M tr\left[\left(\textbf{H}_m\textbf{H}_m^H+\frac{KM \sigma^2_n}{\mathcal{P}_T}\textbf{I}_N\right)^{-2}\textbf{H}_m\textbf{H}^H_m\right]
\end{eqnarray}
which yields result (\ref{soln1}).

\subsection{Passive Beamforming Optimization}
Recall that the phase shifter matrix is a function of the stub length vector, thus the sub-optimization problem for the stub length vector $\bf{\ell}$ can be stated as:
\begin{equation}\label{second:Main}
    \min \limits_{\bf{\ell}} \sum_{m=1}^M\|\alpha \textbf{H}^H_m \textbf{W}_m-\textbf{I}_K\|^2_F
\end{equation}
~~~~~~~~~~~~~~~~~s.t.
\begin{eqnarray}\label{first:a3}
\textbf{H}_m&=&\left[\textbf{H}^H_{su,m}(\mathbf{\Upsilon}^{-1}_m-\textbf{S}_m)^{-1}\textbf{H}_{bs,m}+\textbf{H}^H_{bu,m}\right]^H,\nonumber \\ 
\mbox{diag}(\mathbf{\Upsilon}_m)&=&[v_{11,m},\dots,v_{LL,m}],\nonumber \\ 
v_{pp,m}&=&e^{-2j\beta_m \ell_p},~~~0\leq\ell_p\leq\ell_{\mbox{max}},\nonumber \\ 
v_{pq,m}&=&0 \mbox{ for } p\neq q.\nonumber
\end{eqnarray}

The sub-optimization problem is clearly non-convex. We use gradient descent (GD) algorithm to optimize for $\boldsymbol{\ell}$.
Let us denote the objective function by $G_T=\sum_{m=1}^M G_m$ where $G_m=\|\alpha \textbf{H}^H_m \textbf{W}_m-\textbf{I}\|^2_F$. In the following section, we omit the carrier index $m$ for simplicity. Thus,
\begin{eqnarray}
    G&=&\|\alpha \textbf{H}^H \textbf{W}-\textbf{I}\|^2_F \\ \nonumber
        &=&tr\left[\left(\alpha \textbf{H}_{su}^H\left(\mathbf{\Upsilon}^{-1}-\textbf{S}\right)^{-1} \textbf{H}_{bs}\textbf{W}+\alpha \textbf{H}_{bu}^H \textbf{W}-\textbf{I}\right)\right.\\ \nonumber
    && \left.\left(\alpha \textbf{H}_{su}^H\left(\mathbf{\Upsilon}^{-1}-\textbf{S}\right)^{-1} \textbf{H}_{bs}\textbf{W}+\alpha \textbf{H}_{bu}^H \textbf{W}-\textbf{I}\right)^H\right].\\ \nonumber
    \end{eqnarray}
Thus using complex derivatives and chain rule,
\begin{equation}\label{eq21}
\begin{split}
    \nabla_{\mathbf{\Upsilon}}G=&\mathbf{\Upsilon}^{-T}\left(\mathbf{\Upsilon}^{-1}-\textbf{S}\right)^{-T}\left(\alpha \textbf{H}_{su}^H\right)^T \times \\ \nonumber 
    & \left(\alpha \textbf{H}_{su}^H\left(\mathbf{\Upsilon}^{-1}-\textbf{S}\right)^{-1}\textbf{H}_{bs}W+\alpha \textbf{H}_{bu}^H \textbf{W}-\textbf{I}\right)^* \times \\\nonumber &~~~~~~~~~~~\left(\textbf{H}_{bs} \textbf{W}\right)^T \left(\mathbf{\Upsilon}^{-1}-\textbf{S}\right)^{-T}\mathbf{\Upsilon}^{-T}.
    \end{split}
\end{equation}

We can find
\begin{equation}
   \frac{\partial v_p}{\partial \ell_p}=-2j\beta e^{-2j\beta \ell_p}.
\end{equation}
Then, 
\begin{equation}
    \frac{\partial G}{\partial \ell_p}=2\Re\left\{\left(\nabla_{\mathbf{\Upsilon}}G\right)_p \times\frac{\partial v_p}{\partial \ell_p} \right\}, \; \text{and}
\end{equation}
\begin{equation}
    \nabla_{\boldsymbol{\ell}} \left(G_{T}\right)=\left[\sum_{m=1}^M\frac{\partial G_m}{\partial \ell_1}, \dots, \sum_{m=1}^M\frac{\partial G_m}{\partial \ell_L}\right].
\end{equation}
Moreover, we calculate the update for the stub length based on the rule,
\begin{equation}
    \boldsymbol{\ell}_{t+1}=\boldsymbol{\ell}_t-\mu\left[\nabla_{\boldsymbol{\ell}} \left(G_{T}\right)\right]_t.
\end{equation}
Furthermore, to reduce the likelihood of converging to a local minimum, we adopt two measures. First, we run independent passive beamforming optimizations in parallel from multiple starting points and keep the solution with the highest sum-rate. Second, since the non-convex term $(\boldsymbol{\Upsilon}^{-1}-\mathbf{S})^{-1}$ makes large updates of $\boldsymbol{\ell}$ cause oscillations in the sum-rate, we limit the magnitude of each element's update to a maximum threshold $\xi$.
\par
The overall alternating optimization algorithm is shown in \textbf{Algorithm \ref{Algo1}}. For clarity of presentation, we here illustrate the optimization procedure for a single selected initialization. Note that allocating all power to one carrier and setting $\boldsymbol{\ell}$ to its single-carrier optimum is a feasible point of \eqref{first:main}, so the proposed formulation contains the single-carrier design of \cite{Wij24} as a worst case.

\begin{algorithm}
\caption{Joint optimization algorithm.}\label{Algo1}
\begin{algorithmic}[1]
\State Inputs: $N, K, L, M, a, \mathcal{P}_T,\{f_m\}_{m=1}^M, \mu, \xi, \ell_{\max}$,  \Statex ~~~~~~~~~stopping rule threshold $\Xi$,
\Statex ~~~~~~~~~$\{H_{bs,m}\}_{m=1}^M,\{H_{su,m}\}_{m=1}^M,\{H_{bu,m}\}_{m=1}^M, \sigma_n$
\State Initialize $\boldsymbol{\ell}^{(t=0)}$
\State Calculate $\{S_m\}_{m=1}^M$, $\{\Upsilon_m\}_{m=1}^M$ and $\{H_m\}_{m=1}^M$
\State Calculate $\alpha$ and $\{W_m\}_{m=1}^M$

\Repeat
\State Calculate $\nabla_{\boldsymbol{\ell}} \left(G_{T}\right)$
\For{$p = 1:L$}
\State $\hat{\ell}^{t+1}_{p}={\ell}^t_{p}-\min\left(\max\left(\mu\nabla_{\boldsymbol{\ell}} \left(G_{T}\right)_p,-\xi\right),\xi\right)$
\State ${\ell}^{(t+1)}_{p} = \min\left(\max\left(\hat{\ell}^{(t+1)}_{p},0\right),\ell_{\max}\right)$
\EndFor
\State Calculate $\{\Upsilon_m\}_{m=1}^M$ and $\{H_m\}_{m=1}^M$
\State Calculate $\alpha$ and $\{W_m\}_{m=1}^M$
\State Calculate sum square error $E^{(t+1)}$
\Until{$\frac{E^{(t+1)}-E^{(t)}}{E^{(t)}}<\Xi$}.
\end{algorithmic}
\end{algorithm}

\subsection{Discrete Stub Lengths}
For discrete stub lengths, the elements of $\boldsymbol{\ell}_{t+1}$ are mapped to their nearest valid stub lengths, after determining the continuous optimum stub length values. Here also we make two interesting observations. 
\begin{enumerate}
\item For $q$ discrete stub lengths, uniformly spaced or Lloyd–Max-optimized non-uniform sets \cite{Max60} yield nearly identical results, as the optimized continuous stub lengths are approximately uniformly distributed over the feasible range.
\item Optimizing over continuous stub lengths and quantizing only after the full convergence can yield a lower sum-rate than projecting onto the discrete set at each iteration.
\end{enumerate}

Therefore, we adopt uniformly spaced discrete stub lengths, owing to the simplicity of this approach and its comparable sum-rate performance and perform the projection from the continuous to the discrete set after each iteration.

\subsection{Low-complexity Implementation}\label{lowcomplex}
The aforementioned GD-based optimization requires many iterations due to its constant step size, while multi-start further increases complexity. As joint optimization must be repeated with each channel state information update, often occurring within tens of milliseconds, conventional GD becomes unsuitable for online implementation.\par

To address this, we propose a \textbf{deep unfolding framework} with a fixed number of $T$ iterations and offline-learned GD step sizes. Each iteration forms a network layer with a trainable step size $\mu_t$, optimized using Adam and backpropagation across all $T$ layers over diverse channel realizations and initializations (details omitted for brevity). During online operation, the learned step sizes eliminate line searches and multiple initializations, significantly reducing complexity compared with multi-start GD in Section \ref{jointOpt} and enabling real-time implementation.

\section{Simulation Results and Discussion}
We evaluate the performance of the proposed multi-carrier RIS-aided MIMO communication system together with the joint beamforming optimization framework. We consider a communication system with the following parameters: $N=8, K=8, L=64, \sigma_n^2=-100$ dBm. We also assume the channels to be complex Gaussian with zero mean and $0.5$ variance per dimension, together with a path loss. Path loss exponents for BS to RIS, BS to users and RIS to users are $2.5, 4.5, 4.5$ respectively. BS to RIS distance is $500$ m and the mean distance from BS to users and RIS to users are $500$ m and $100$ m respectively. Moreover, we consider carriers centered around $5$ GHz so that $\lambda_{\mbox{mean}}=60$ mm and $a=\lambda_{\mbox{mean}}/4$. Carrier spacing is $100$ MHz. For the optimization algorithm we select $\xi=5\times 10^{-5}, \Xi=5\times 10^{-6}$, number of multi-start points $=50$ and $\ell_{\mbox{max}}=30$ mm. All simulations were performed in MATLAB R2024 on a computer equipped with a 2.4 GHz Intel Core i5 processor and 8 GB of RAM.S

\par
\begin{figure*}[!t]
    \centering
    \begin{minipage}{0.32\textwidth}
        \centering
        \includegraphics[width=1.04\textwidth]{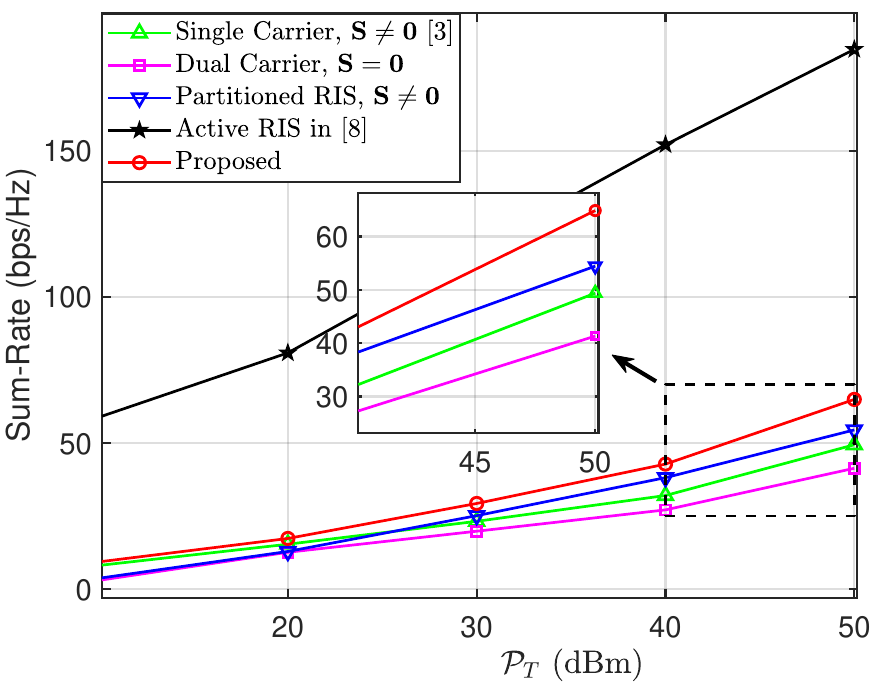}
        \vskip 0.2cm
        \caption{Sum-rate variation with BS power.}
        \label{MainComparefig}
    \end{minipage}
    \hskip 0.2cm
    \begin{minipage}{0.32\textwidth}
        \centering
        \includegraphics[width=1\textwidth]{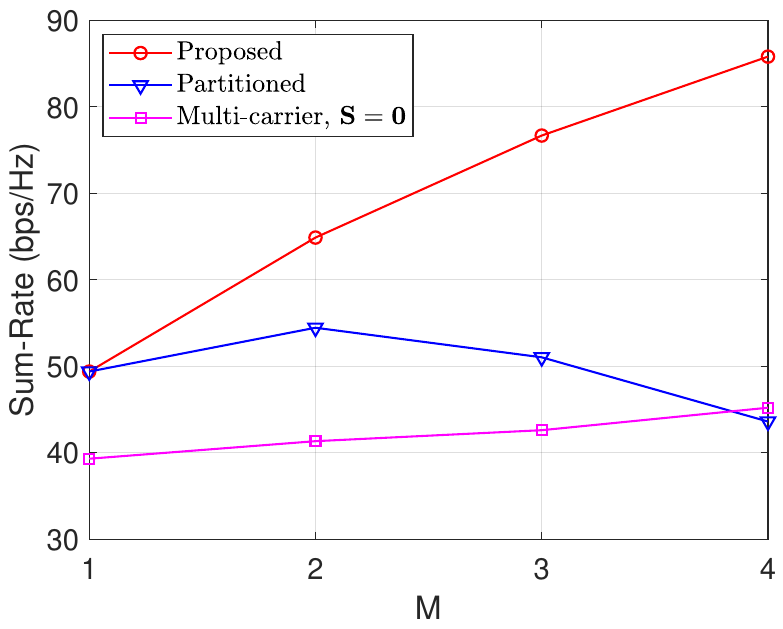}
        \vskip 0.2cm
        \caption{Sum-rate variation with number of carriers, $\mathcal{P}_T=50$ dBm.}
        \label{varWithMfig}
    \end{minipage}
    \hskip 0.2cm
     \begin{minipage}{0.32\textwidth}
        \centering
        \includegraphics[width=1\textwidth]{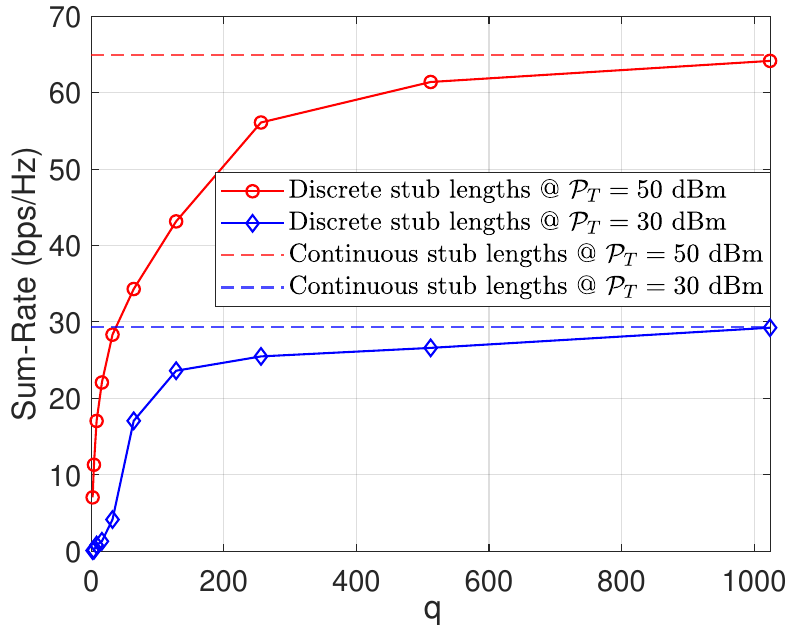}
        \vskip 0.2cm
        \caption{Sum-rate variation with number of discrete stub length levels, $M=2$.}
        \label{DescreteStubsfig}
    \end{minipage}
\end{figure*}

Optimization employs either the physically consistent or conventional ($\mathbf{S}=\mathbf{0}$) model, depending on the scenario, while sum-rate is always evaluated using the former to reflect practical conditions. Fig. \ref{MainComparefig} compares the proposed dual-carrier model and joint optimization with prior works, showing nearly $20\%$ improvement over partitioned RIS at $\mathcal{P}_T=50$ dBm. Although dual carrier partitioning enables carrier-specific phase optimization, fewer effective RIS elements per carrier outweigh this benefit compared with simultaneous proposed dual-carrier operation. The proposed system also outperforms the physically consistent single-carrier approach \cite{Wij24} and conventionally optimized dual-carrier system, which lack multiplexing and physically consistent optimization gains, respectively. While the active RIS approach \cite{Tua24} with 2 carriers achieves higher sum-rates, it incurs greater power consumption and circuit complexity. Thus, the proposed system offers a favorable performance–complexity trade-off.\par

Fig. \ref{varWithMfig} compares sum-rate variations with the number of carriers, $M$. The proposed approach achieves increasing sum-rates with $M$, whereas partitioned RIS performance eventually declines due to fewer RIS elements per carrier. Although conventional multi-carrier optimization also improves with $M$, it remains inferior to the proposed approach due to the absence of physically consistent optimization benefits.\par

Fig. \ref{DescreteStubsfig} illustrates the trade-off between stub-length resolution and sum-rate, with approximately 250 levels approaching the continuous-stub benchmark. In practical implementations, PIN-diode-switched stubs offer lower power consumption but a larger form factor, whereas adjustable stubs provide a more compact alternative with higher power consumption and slower response time. Thus, the choice of implementation depends on the required trade-off among sum-rate performance, power consumption, size, and response speed.\par

\begin{figure}[!thb]
    \centering
    \includegraphics[width=0.33\textwidth]{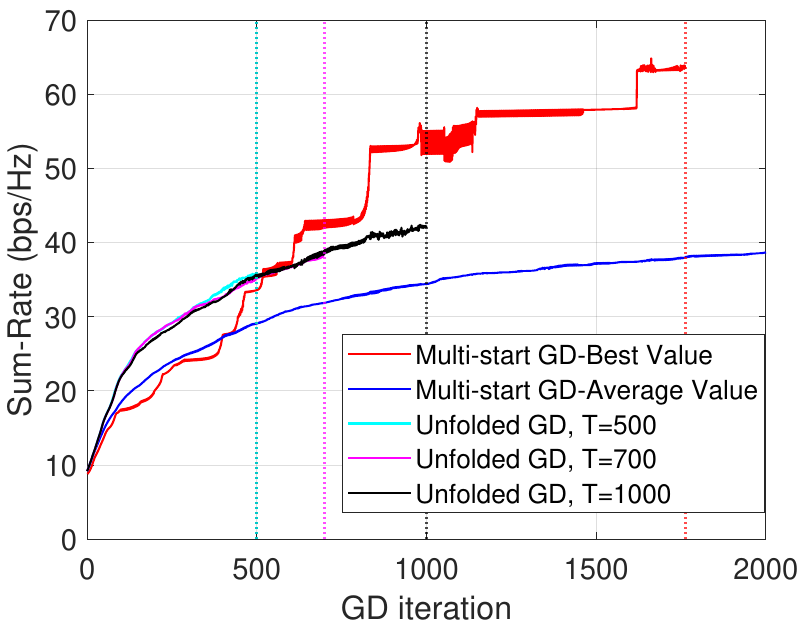}
    \vskip 0.2cm
    \caption{Convergence of the optimization algorithm, $M=2, \mathcal{P}_T=50$ dBm.}
    \label{unfoldingfig}
\end{figure}

Fig.~\ref{unfoldingfig} compares the convergence of multi-start GD with constant $\mu$ and unfolded GD for $M=2$ and $\mathcal{P}_T=50$~dBm. After 2000 iterations, the best multi-start run achieves $64$~bps/Hz versus $40$~bps/Hz on average, at a substantial runtime cost. Unfolded GD, with step sizes learned via offline training and $T=700$, surpasses the multi-start average, from a single start. On the same computing platform, multi-start GD with 50 initializations takes $114.2$~s, whereas unfolded GD takes only $0.96$~s online, corresponding to a 119-fold reduction in execution time. Therefore, unfolding offers a computationally-efficient alternative to optimize system performance.\par

\section{Conclusion}\label{conc}
We have proposed a physically-consistent joint beamforming framework for RIS-aided multi-carrier MIMO systems, enabling simultaneous multi-carrier operation using all RIS elements. By exploiting multiple reflections, the proposed approach outperforms conventional and partitioned RIS schemes, achieving nearly $20\%$ sum-rate improvement over partitioning. We further investigated practical stub-length implementations and developed a deep unfolding approach that reduces online optimization complexity. 


\begin{thebibliography}{7}
\bibitem{Has24} S. Hasan, et al., ``A Comprehensive Review on Reconfigurable Intelligent Surface for 6G Communications: Overview, Deployment, Control Mechanism, Application, Challenges, and Opportunities," {\em Wireless Pers. Commun.}, vol.  139, pp. 375–429, Oct 2024.

\bibitem{Daj22} M. Dajer et al., ``Reconfigurable intelligent surface: Design the
channel-A new opportunity for future wireless networks," {\em Digit. Commun. Netw.}, vol. 8, no. 2, pp. 87–104, Apr. 2022.

\bibitem{Wij24} D. Wijekoon, A. Mezghani, and E. Hossain, ``Phase Shifter Optimization in RIS-Aided MIMO Systems Under Multiple Reflections," {\em IEEE Trans. Wireless Commun.}, vol. 23, no. 8, pp. 8969-8983, Aug. 2024.

\bibitem{Che20} W. Chen et al., ``Angle-Dependent Phase Shifter Model for Reconfigurable Intelligent Surfaces: Does the Angle-Reciprocity Hold?," {\em IEEE Commun. Lett.}, vol. 24, no. 9, pp. 2060-2064, Sept. 2020.

\bibitem{Cao24} L. Cao, H. Yin, L. Tan, and X. Pei, ``RIS With Insufficient Phase Shifting Capability: Modeling, Beamforming, and Experimental Validations," {\em IEEE Trans. Commun.}, vol. 72, no. 9, pp. 5911-5923, Sept. 2024.


\bibitem{Cai22} W. Cai, et al., ``IRS-Assisted Multicell Multiband Systems: Practical Reflection Model and Joint Beamforming Design," {\em IEEE Trans. Commun.}, vol. 70, no. 6, pp. 3897-3911, June 2022.

\bibitem{Sol23} M. Soleymani, I. Santamaria, A. Sezgin, and E. Jorswieck, ``Maximization of Minimum Rate in MIMO OFDM RIS-Assisted Broadcast Channels," in Proc. {\em IEEE 9th Int. Workshop Comput. Adv. Multi-Sensor Adapt. Process. (CAMSAP)}, Herradura, Costa Rica, 2023, pp. 11-15.

\bibitem{Tua24} H. D. Tuan, A. A. Nasir, E. Dutkiewicz, H. V. Poor, and L. Hanzo, ``Active-RIS Enhances the Multi-User Rate of Multi-Carrier Communications," {\em IEEE Trans. Veh. Technol.}, vol. 73, no. 11, pp. 16948-16963, Nov. 2024.

\bibitem{Mez22} A. Mezghani, F. Bellili, and E. Hossain, ``Reconfigurable Intelligent Surfaces for Quasi-Passive mmWave and THz Networks: Should They be Reflective or Redirective?," in Proc. {\em 56th Asilomar Conf. Signals, Syst., Comput.}, Pacific Grove, CA, USA, 2022, pp. 1076-1080.

\bibitem{Reh21} H. Ur Rehman, F. Bellili, A. Mezghani and E. Hossain, "Joint Active and Passive Beamforming Design for IRS-Assisted Multi-User MIMO Systems: A VAMP-Based Approach," in {\em IEEE Trans. Commun.}, vol. 69, no. 10, pp. 6734-6749, Oct. 2021.


\bibitem{Max60} J. Max, ``Quantizing for minimum distortion," {\em IRE Trans. Inf. Theory}, vol. 6, no. 1, pp. 7-12, March 1960.
\end{thebibliography}
\end{document}